\documentclass[aps,prl,reprint,showpacs,superscriptaddress]{revtex4-1}

\usepackage{amsmath}
\usepackage{xspace} 
\usepackage{graphicx}
\usepackage{natbib}
\usepackage{hyperref}
\usepackage[utf8]{inputenc}
\hypersetup{
	colorlinks   = true, 
	urlcolor     = blue, 
	linkcolor    = black,
	citecolor   = black 
}

\newcommand{\nm}{\,\ensuremath{{\rm nm}}\xspace}

\newcommand{\GHz}{\,\ensuremath{{\rm GHz}}\xspace}
\newcommand{\MHz}{\,\ensuremath{{\rm MHz}}\xspace}
\newcommand{\kHz}{\,\ensuremath{{\rm kHz}}\xspace}

\newcommand{\GPa}{\,\ensuremath{{\rm GPa}}\xspace}

\newcommand{\mbar}{\,\ensuremath{{\rm mbar}}\xspace}

\newcommand{\mW}{\,\ensuremath{{\rm mW}}\xspace}

\newcommand{\fg}{\,\ensuremath{{\rm fg}}\xspace}

\newcommand{\degree}{\ensuremath{{^\circ}}\xspace}

\begin{document}
	
	\title[]{GHz Rotation of an Optically Trapped Nanoparticle  in Vacuum}
	
	\author{Ren\'{e}~\surname{Reimann}}
	\email{rreimann@ethz.ch}
	\author{Michael~\surname{Doderer}}
    \author{Erik~\surname{Hebestreit}}
	\author{Rozenn~\surname{Diehl}}
	\author{Martin~\surname{Frimmer}}
	\author{Dominik~\surname{Windey}}
	\author{Felix~\surname{Tebbenjohanns}}
	\author{Lukas~\surname{Novotny}}

	\affiliation{Photonics Laboratory, ETH Z{\"u}rich, 8093 Z{\"u}rich, Switzerland}
	
	
	\begin{abstract}
	We report on rotating an optically trapped silica nanoparticle in vacuum by transferring spin angular momentum of light to the particle's mechanical angular momentum. At sufficiently low damping, realized at pressures below $10^{-5}\mbar$, we observe rotation frequencies of single $100\nm$ particles exceeding $1\GHz$. We find that the steady-state rotation frequency scales linearly with the optical trapping power and inversely with pressure, consistent with theoretical considerations based on conservation of angular momentum. Rapidly changing the polarization of the trapping light allows us to extract the pressure-dependent response time of the particle's rotational degree of freedom. 
	\end{abstract}

	\maketitle

	\paragraph*{Introduction.} Optomechanics is the science of measuring and controlling mechanical motion using light~\cite{Aspelmeyer2014}. One particularly interesting optomechanical system is a dielectric nanoparticle levitated in a strongly focused laser beam using the forces of light~\cite{Ashkin2007, Chang2010,Romero-Isart2011, Li2011c, Gieseler2012,Yin2013}.
	The trapping laser confines the particle to the focal region and scatters off the particle, providing a measurement of its center-of-mass motion. Using active feedback mechanisms and autonomous cavity-assisted cooling schemes, the center-of-mass motion of an optically levitated nanoparticle has been controlled to a remarkable degree, putting the quantum regime of mechanical motion within reach~\cite{Kiesel2013,Millen2015,Jain2016,Vovrosh2017}. 
	Recently, researchers have started to turn their attention to rotational degrees of freedom of levitated objects~\cite{Shi2016}. Taking inspiration from optically induced rotation of particles trapped in liquid media~\cite{Friese1998}, the torsional and rotational motion of optically levitated objects featuring shape asymmetries or anisotropic optical properties have been investigated~\cite{Arita2011,Arita2013,Hoang2016,Kuhn2017,Kuhn2017a}.
	Gaining control over the rotation of a levitated object is interesting from two perspectives. First, in high vacuum, optically levitated particles offer the potential to reach extremely high rotation speeds~\cite{Kane2010,Nagornykh2017,Kuhn2015}, necessary to investigate unexplored types of rotation-induced fluctuating forces and vacuum-friction effects~\cite{Zhao2012,Manjavacas2017}.
	Second, adding rotational control to the toolbox of levitated optomechanics is appealing when considering the regime of low excitation numbers currently investigated for the translational degrees of freedom. Each center-of-mass degree of freedom of a trapped particle embodies a quantum mechanical harmonic oscillator with equidistantly spaced energy levels and finite ground-state energy~\cite{Aspelmeyer2014}. In contrast, the rotational degrees of freedom offer a nonlinear energy spectrum with vanishing ground-state energy~\cite{Shi2016}.
	Accessing this new regime of rich mesoscopic physics requires optical control of both the rotational and the center-of-mass motion in high vacuum.
	Importantly, while the center-of-mass motion of silica nanoparticles can be cooled close to the quantum regime~\cite{Jain2016}, controlling the rotational motion of such a particle  in high vacuum has remained elusive to date. 
	
	In this Letter, we measure the rotational motion of an optically trapped silica nanoparticle with a diameter of 100~nm. We drive the particle's rotation using circularly polarized light. At pressures below $10^{-5}\mbar$, we reach rotation frequencies exceeding $1\GHz$. To our knowledge, these are the highest rotation frequencies of a mechanical object that have been realized to date.
	\paragraph*{Experimental setup.}
	\begin{figure}[b!]
		\centering{\includegraphics[width=1\columnwidth]{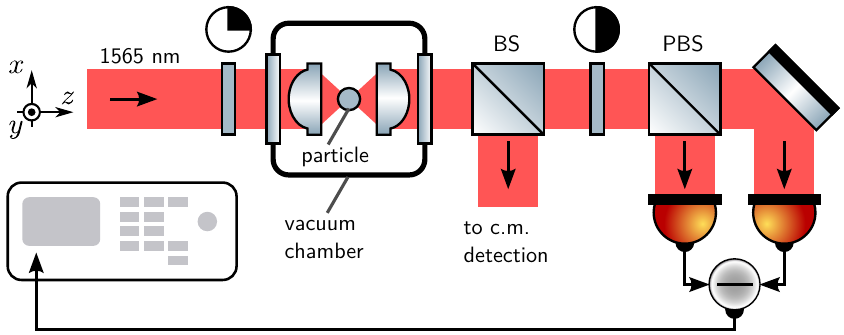}}\\
		\caption{Simplified experimental setup. The polarization state of an initially linearly polarized laser beam can be set by a quarter-wave plate before entering a vacuum chamber. Inside the chamber, an optical trap for a nanoparticle is formed by focusing the beam with an aspheric lens (0.77 NA). The light is collected by an identical lens and equally split by a beam splitter (BS). One half of the power is utilized for center-of-mass (c.m.) detection of the particle. The other half is sent onto a half-wave plate followed by a polarizing beam splitter (PBS), which enables balanced detection of the particle rotation via a spectrum analyzer. }
		\label{Fig.setup}
	\end{figure}
	Our experimental setup is depicted in Fig.~\ref{Fig.setup}. A laser beam (wavelength $\lambda = 1565\nm$, linearly polarized along the $x$ axis) propagates along the $z$ direction. Before entering a vacuum chamber, the polarization of the light can be set from linear over elliptical to circular by means of a quarter-wave plate. An aspheric lens (0.77 NA) inside the chamber focuses the beam to a diffraction-limited spot that forms an optical tweezer trap for a single silica nanoparticle with a nominal diameter of $100\nm$. An identical lens collects and collimates the light for detection. 
	A nonpolarizing beam splitter (BS) behind the vacuum chamber splits the beam such that half of the optical power is used for detecting the particle's center-of-mass motion, as described in Ref.~\cite{Gieseler2012}, and the other half for measuring its rotation. 
	The center-of-mass motion of the particle shows three distinct oscillation frequencies, corresponding to the motion of the particle along the $x$, $y$, and $z$ direction. The transverse center-of-mass oscillation frequencies $\Omega_{\text{c.m.}}^{(x)}$ and $\Omega_{\text{c.m.}}^{(y)}$ along the $x$ and $y$ direction are around $2\pi\times100\kHz$ and sensitively depend on the polarization of the trapping light. This dependence can be utilized to cross-check the polarization of the trapping light, as $\Omega_{\text{c.m}}^{(x)}$ and $\Omega_{\text{c.m.}}^{(y)}$ are maximally distinct for linear polarization and become degenerate for circular polarization. 
	In order to measure the particle's rotation frequency $\Omega_\mathrm{rot}$, the second half of the detection beam passes a half-wave plate before it is split at a polarizing beam splitter (PBS) and sent onto a fast balanced photodetector (bandwidth $1.6\GHz$), which is connected to a spectrum analyzer (bandwidth $20\GHz$) \cite{Arita2013, Hoang2016}.
	Independent of the rotation detection mechanism---which can arise from residual birefringence~\cite{Garetz1979, Arita2013}, from asymmetric particle shape~\cite{Kuhn2017a}, or the angular Doppler shift~\cite{Garetz1981}---we expect the signal at the spectrum analyzer to oscillate at $2\Omega_\mathrm{rot}$.       
	
\begin{figure}[t!]
		\centering{\includegraphics{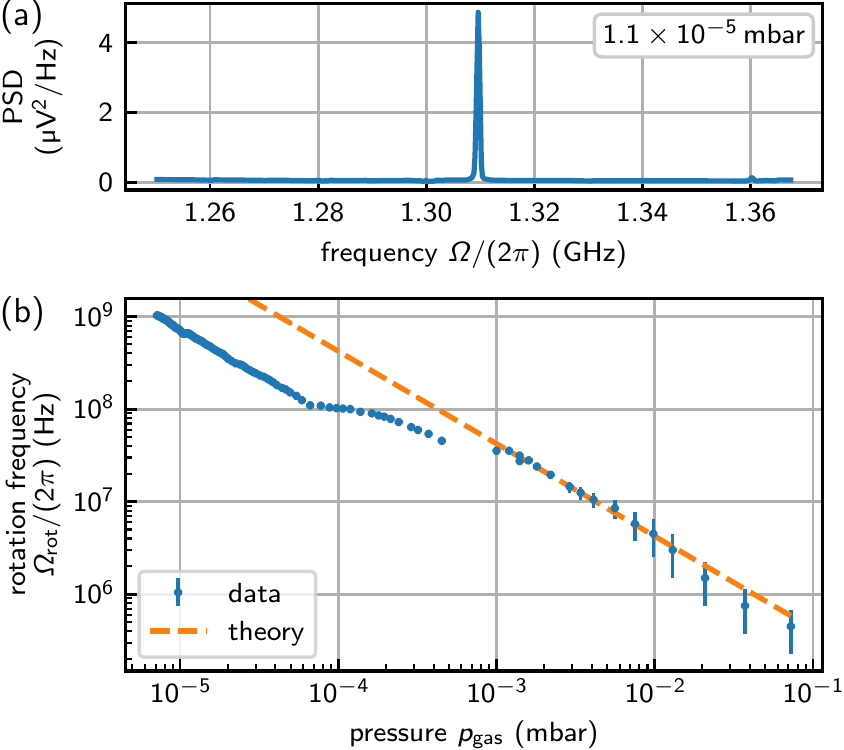}}\\
		\caption{Rotation frequency as a function of pressure at a focal trapping power of $226(5)\mW$ and for nearly left-circularly polarized (LCP) trapping light. (a)~Measured power spectral density (PSD) of the rotation signal at a pressure of $1.1\times 10^{-5}\mbar$ showing a signal at $1.31\GHz$, which corresponds to a rotation frequency of $\Omega_\mathrm{rot}/(2\pi)=655\,\mathrm{MHz}$. (b)~Measured rotation frequency for varying gas pressure with a maximum rotation frequency of $\Omega_\mathrm{rot}/(2\pi)=1.029(1)\,\mathrm{GHz}$ at a pressure of $7.2\times10^{-6}\mbar$. 
		We attribute the deviation between theory and experiment below $10^{-3}\mbar$ to an underestimation of pressure by the used ion gauge.}
		\label{Fig.pressure_rot}
	\end{figure}

	\paragraph*{Results and discussion.} For our first set of measurements, we adjust the polarization of the trapping laser close to circular. In Fig.~\ref{Fig.pressure_rot}(a), we show a typical rotation spectrum, recorded at a pressure of $10^{-5}\mbar$ and a trapping laser power $P=226(5)~$mW. The signal at $1.31\GHz$ corresponds to a particle rotation frequency $\Omega_\mathrm{rot}/(2\pi)=655\MHz$. In Fig.~\ref{Fig.pressure_rot}(b), we plot the observed rotation frequency $\Omega_\text{rot}$ as a function of pressure in the vacuum chamber. As the pressure is decreased from $10^{-1}\mbar$ to below $10^{-5}\mbar$, $\Omega_\mathrm{rot}/(2\pi)$ increases from a few hundred $\kHz$ to above $1\GHz$, where our measurement is currently limited by the finite bandwidth of our photodetector. 
	
	To understand the pressure dependence of $\Omega_\mathrm{rot}$ observed in Fig.~\ref{Fig.pressure_rot}(b), we consider the equation of motion of the particle's angular momentum $L$, whose time rate of change equals the sum of all applied torques
	\begin{equation}
	\label{Eq.ang_momentum}
	\frac{d}{dt}L   = I \frac{d}{dt}\Omega_{\text{rot}} =  \tau_{\text{opt}} + \tau_{\text{drag}}. 
	\end{equation}
	We approximate the particle as a sphere with a moment of inertia $I=0.4 m R^2$, mass $m\approx 1\fg$, and radius $R= 50\nm$. The optical torque $\tau_\text{opt}$ arises from the particle's interaction with the laser field. The drag $\tau_\text{drag}$ is due to the interaction of the particle with the residual gas in the vacuum chamber. We consider three possible contributions to the optical torque $\tau_{\text{opt}} = \tau_{\text{abs}} + \tau_{\text{brf}} + \tau_{\text{shape}}$. The first component of the optical torque $\tau_{\text{abs}}=\sigma_{\text{abs}} \Delta s_{\text{abs}} \mathcal{I} \lambda / (2\pi c)$ originates from absorbed photons that transfer their spin angular momentum to the particle~\cite{Arita2011}. Here, $\sigma_{\text{abs}}$ is the particle's absorption cross section, $\mathcal{I}$ is the intensity at the particle's position, and $c$ is the speed of light. 
	The degree of circular polarization $\Delta s_{\text{abs}} \in [-1,1] $ becomes $-1$ for left-circularly polarized (LCP) trapping light and $1$ for a right-circularly polarized (RCP) field. The second optical torque component $\tau_{\text{brf}}\propto \mathcal{I}$ arises from photons changing their polarization state when scattering off the particle. This torque exists only for a particle exhibiting a finite birefringence and depends on the difference between the ordinary and the extraordinary refractive indices of the particle material~\cite{Friese1998}. The third optical torque component $\tau_{\text{shape}}\propto \mathcal{I}$ describes the force arising due to a possible shape asymmetry of the particle~\cite{Kuhn2017a}. 
	On the other hand, the torque $\tau_{\text{drag}}$ in Eq.~\eqref{Eq.ang_momentum} damps the particle's rotation due to viscous interaction with gas molecules in the vacuum chamber. Our experiment operates in a regime where the mean free path of gas molecules is much bigger than the particle diameter. In this regime, the viscous torque is given by $\tau_{\text{drag}} = - I \Omega_{\text{rot}}/t_{\text{damp}}$~\cite{Fremerey1982}. The damping time $t_{\text{damp}} = \beta m \overline{v} / (p_{\text{gas}} R^2)$ depends on the pressure $p_{\text{gas}}$, on the mean molecular velocity $\overline{v}$, and on an accommodation factor $\beta$ that takes into account the efficiency with which molecules transfer angular momentum via collisions to the particle.
	
	In the regime of almost circularly polarized light, any torque contribution trying to align the particle relative to the polarization ellipse of the trapping field can be neglected. 
	Therefore, the particle rotates continuously at a frequency $\Omega_{\text{rot}} \propto \tau_\text{opt}/p_{\text{gas}}$, which can be determined by solving for the steady state of Eq.~\eqref{Fig.setup}. The inverse scaling of the rotation frequency with pressure is displayed as the dashed line in Fig.~\ref{Fig.pressure_rot}(b). 
	We interpret the deviation of the measured data from the model as an artifact known as gauge pumping~\cite{StanfordResearchSystems,Winters1962}. This effect arises for ion gauges (as used in our experiment at pressures below $10^{-3}\mbar$) and leads to an underestimation of the pressure at the particle position.
	
	\begin{figure}[t!]
		\centering{\includegraphics{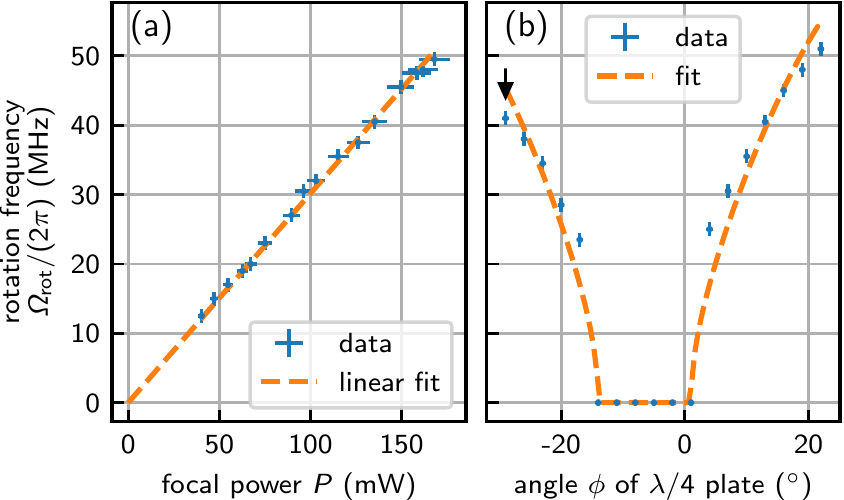}}\\
		\caption{(a)~Power dependence of the rotation frequency at a pressure $p_{\text{gas}} = 1.0\times10^{-4}\mbar$. The polarization of the trapping light equals the one in Fig.~\ref{Fig.pressure_rot} and corresponds to  a quarter-wave plate angle of $-29\degree$, see arrow in (b). (b)~Particle rotation frequency as a function of quarter-wave plate angle at a focal power of $226(5)\mW$. 
		The wave plate angle determines the degree of linear vs circular polarization of the trapping light. A finite degree of circular polarization is necessary to induce rotation of the particle. The orange dashed line is a fit to $\Omega_\mathrm{rot}=\mathrm{Re}[a\sqrt{(1-\cos b)^2\sin^2(2\phi - \phi_0)-\sin^2(b)\cos^2(2\phi-\phi_0)}]$ \cite{Friese1998}.
		The data in (a) and (b) have been recorded with different particles.}
		\label{Fig.power_rot}
	\end{figure}
	
	Having investigated the pressure dependence of the rotation frequency, we turn to its dependence on the power of the trapping laser. 
	In the limit of nearly circularly polarized trapping light, the optical torque scales linearly with optical power $\tau_\text{opt} \propto P$, such that we find for the particle's rotation frequency $\Omega_\text{rot} \propto P$.
	Keeping the polarization of the trapping light nearly circularly polarized at constant pressure $p_\mathrm{gas}=1\times10^{-4}~\text{mbar}$, we measure the rotation frequency $\Omega_\text{rot}$ as a function of focal power $P$. The result is displayed in Fig.~\ref{Fig.power_rot}(a) and shows very good agreement with the theoretically expected linear scaling. 
	
	Thus far, our experiments have been carried out with nearly circularly polarized trapping light. In Fig.~\ref{Fig.power_rot}(b), we investigate the dependence of the particle's rotation frequency on the polarization state of the light field by varying the angle $\phi$ between the fast axis of the quarter-wave plate in front of the vacuum chamber and the polarization vector of the initially linearly polarized light (see setup in Fig.~\ref{Fig.setup}). 
	We observe a vanishing rotation frequency for angles $\phi$ between $-14^\circ$ and $0^\circ$ and an increasing rotation frequency with increasing (absolute) value of $\phi$ outside that range. 
	This experimental result can be explained by remembering that both $\tau_\text{brf}$ and $\tau_\text{shape}$ have two contributions~\cite{Friese1998, Kuhn2017a}. The first contribution leads to a restoring torque which tends to align the particle's symmetry axes (given by its shape or birefringence) to the main axes of the polarization ellipse. This contribution vanishes for perfectly circularly polarized light and is maximized for linearly polarized light. The second contribution drives the particle rotation. It vanishes for a linearly polarized field and increases with the degree of circular polarization.
	We conclude that for angles between $-14\degree$ and $0\degree$ the trapping light is predominantly linearly polarized, which leads to a dominant restoring torque pinning the particle's orientation to the polarization ellipse. As the quarter-wave plate is rotated beyond that range, the torque contribution leading to rotation overcomes that pinning the particle's orientation and the particle starts to rotate.
	Finally, we turn to the observation that the data in Fig.~\ref{Fig.power_rot}(b) are not symmetric around $\phi=0$, as expected. We attribute the horizontal shift of the data by roughly $-6^\circ$, as well as the slight deviation from the fit, to the birefringence of the vacuum window and the aspheric trapping lens.
	We also note that for extremely pure LCP or RCP light, the particle quickly escapes from the trap.
	We speculate that this instability might be due to a spin-orbit coupling of the particle's rotation and its center-of-mass motion, similar to the effects observed in Ref.~\cite{Arita2013}.
	
\begin{figure}[t]
		\centering{\includegraphics{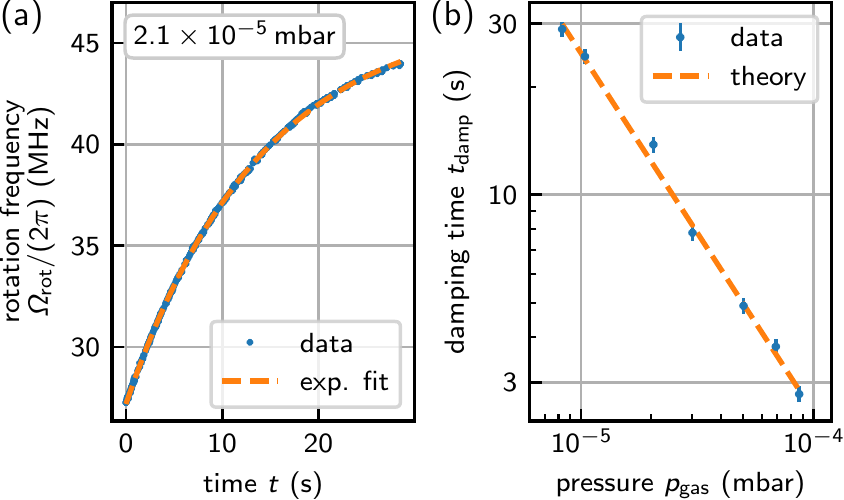}}\\
		\caption{Measurement of system dynamics at a focal power $P=230(5)\mW$. (a) After a steplike change of the optical torque by a fast rotation of the quarter-wave plate [see Fig.~\ref{Fig.setup} and Fig.~\ref{Fig.power_rot}(b)] we measure the rotation frequency as a function of time at a fixed pressure. A fit (dashed line) of $\Omega_{\text{rot}} = c_1 + c_2 \exp(-t/t_{\text{damp}})$, with $t_\text{damp}$, $c_1$, and $c_2$ as free parameters, reveals the characteristic response time $t_\text{damp}$ of our system. (b) Repeating the procedure described in (a) for different pressures results in a damping time which scales as $t_\text{damp}\propto 1/p_\text{gas}$ (dashed line). }
		\label{Fig.damping_time}
	\end{figure}

	In a final experiment, we study the timescale of equilibration of the rotational dynamics of the levitated particle. To this end, we apply a steplike increase of the optical torque applied to the particle.  
	Experimentally, we rapidly change the angle $\phi$ of the quarter-wave plate from $-19\degree$ to $-29\degree$ and record $\Omega_\text{rot}$ as a function of time $t$ at fixed pressure. 
	In Fig.~\ref{Fig.damping_time}(a), we show the time dependence of the observed rotation frequency after switching the polarization state of the trapping light. We observe a nonlinear acceleration of the rotation frequency. We explain our observation by considering the time dependent solution of Eq.~(1), yielding $\Omega_{\text{rot}} = c_1 + c_2 \exp(-t/t_{\text{damp}})$, with constants $c_1$ and $c_2$ that depend on the initial parameters. Indeed, the data in Fig.~\ref{Fig.damping_time}(a) fit well to the expected exponential behavior (dashed line). We repeat this experiment at different pressures, extract $t_\text{damp}$ from the fits, and plot the damping times in Fig.~\ref{Fig.damping_time}(b). As expected, the damping time scales inversely with pressure according to $t_\text{damp} \propto 1/p_\text{gas}$.
	
	We note that the data displayed in the figures of this manuscript have been recorded with different particles. The reason is that particles occasionally escape from the trap or possibly disintegrate due to high rotation frequencies and the associated high centrifugal forces acting on the particle~\cite{Schuck2018}. The properties (including exact diameter, shape, birefringence, absorption, surface roughness) of each particle may therefore vary between figures. Nevertheless, all rotation states reported in this paper have been observed to be stable on a minimal timescale of minutes.
	
	\paragraph*{Conclusion.} We have demonstrated the stable rotation of an optically trapped dielectric particle of $100\nm$ diameter at rotation frequencies exceeding $1\GHz$.  With a simple model, we were able to describe our experimental observations. However, while our model yields the correct scaling of the parameters involved (e.g. laser power and pressure) it does not identify the dominating torque transfer mechanism. Ongoing work is aimed at elucidating this mechanism. 
	
	Our results have important implications for quantum optomechanics, cosmology and material tests at the nanoscale: 
	Together with the fact that the center-of-mass motion of optically levitated nanoparticles can be cooled to the sub-mK regime~\cite{Jain2016}, our control over the rotational degree of freedom could be utilized for studies of friction at a fundamental level~\cite{Zhao2012,Manjavacas2017}. Another exciting prospect is to explore the interaction of center-of-mass and rotational degrees of freedom, which may allow studies of spin-orbit coupling in optically levitated systems. Such a coupling, which has been reported in Ref.~\cite{Arita2013}, is not yet observed in our experiment, for reasons to be investigated.
	Interestingly, the ability to rotate particles at GHz frequencies might provide a platform to study questions arising in the context of cosmology. Rapidly spinning charged dust particles in the interstellar medium have been proposed to be responsible for $\GHz$ radiation in measurements of the cosmic background radiation~\cite{Draine1998}. Together with controlling the particle charge~\cite{Frimmer2017}, our method of rotating a nanoparticle at \GHz frequencies could enable a test bed for this hypothesis. 
	In the direction of more applied research~\cite{Schuck2018}, rapidly rotating nanoparticles with circumferential speeds exceeding $300\,\text{m}/\text{s}$ and radial accelerations on the surface of more than $10^{12}\, {\rm m/s^2}$ (corresponding to the gravitational acceleration on the surface of a neutron star~\cite{Green2004}) can be utilized to test material limits under centrifugal stress on the nanoscale. For our glass particles with density $\rho = 2000 \,{\rm kg/m^3}$, the realized maximal tensile strength $\sigma_{\rm tens} \approx \rho \Omega_{\rm rot}^2 R^2$~\cite{Schuck2018} is about $0.2\GPa$. Accordingly, our experiments operate in the interesting regime close to the ultimate tensile strength on the order of $10\GPa$~\cite{LeBourhis2007} at which defect-free glass would disintegrate.
	
	\paragraph*{Acknowledgments.}
	\begin{acknowledgments}
		We thank P. Kurpiers and A. Wallraff for lending us a high bandwidth spectrum analyzer.
		This work has been supported by ERC-QMES (No.~338763). 
		R.R acknowledges funding from the European Union’s Horizon 2020 research and innovation program under the Marie Skłodowska-Curie Grant Agreement No.~702172.\\
		R.R. and M.D. contributed equally to this work.
	\end{acknowledgments}
	
	\paragraph*{Note added.} We have recently become aware of related work on optically rotating micron-sized spheres in high vacuum~\cite{Monteiro2018} and on $\GHz$ rotation of nanodumbbells~\cite{Ahn2018}.
	\bibliographystyle{apsrev4-1}
	\bibliography{rotating_particle_bib}

\end{document}